\documentstyle{mn}
\input{psfig.sty}

\title{Time evolution of the linear perturbations of a rotating Newtonian
polytrope}  

\author[D. I. Jones, N. Andersson and N. Stergioulas]
{D. I. Jones$^1$, N. Andersson$^1$ and N. Stergioulas$^2$\\
$^1$ Faculty of Mathematical Studies, University of Southampton, 
     Highfield, Southampton, SO17 1BJ, United Kingdom \\
$^2$ Department of Physics, Aristotle University of Thessoloniki,
Thessaloniki 54006, Greece \\
}

\begin{document}

\maketitle

\begin{abstract}

We present the results of numerical time evolutions of the linearised
perturbations of rapidly and rigidly rotating Newtonian polytropes while
making the Cowling approximation.  The evolution code runs stably for
hundreds of stellar rotations, allowing us to compare our results with
previously published eigenmode calculations, for instance the f-mode
calculations of Ipser \& Lindblom, and the r-mode calculations of Karino et
al.  The mode frequencies were found to be in agreement within the expected
accuracy.  We have also examined the inertial modes recently computed by
Lockitch \& Friedman, and we were able to extend their slow-rotation
results into the rapid rotation regime.  In the longer term, this code will
provide a platform for studying a number of poorly understood problems in
stellar oscillation theory, such as the effect of differential rotation and
gravitational radiation reaction on normal mode oscillations and, with
suitable modifications, mode-mode coupling in the mildly non-linear regime.

\end{abstract}


\section{Introduction}
\label{sect:ptbs}

The study of the normal modes of stars has long been recognised as a
subject of considerable importance.  The electromagnetic signatures of
such oscillations have been used for decades to probe the structure of
main sequence stars (Cox 1980).  More recently the modes of compact
stars (neutron stars and white dwarfs) have received attention from
theorists, again with a hope of converting observations into
constraints on stellar structure.  Such modes may have electromagnetic
signatures (van Horn 1980), excited perhaps during accretion or in a
thermonuclear burst (Van der Klis 2000).  Equally, an
oscillating compact star is a source of gravitational radiation, and
may even be unstable to such an energy loss via the so-called
Chandrasekhar-Friedman-Schutz mechanism (hereafter CFS; see Friedman
\& Schutz 1978a,b).  Particularly in this last regard, it is rapidly
rotating compact stars that are of most interest, as only in the case
of rapid rotation might gravitational radiation reaction overcome the
dissipative effects of viscosity to amplify a mode.

Very few previous studies have examined modes of rapidly rotating
compact stars in Newtonian theory (see Stergioulas 1998, Andersson \&
Kokkotas 2001 for reviews of relativistic computations).  The f-modes
were calculated by Ipser \& Lindblom (1990), the r-modes by Karino et
al.\ (2000), and the inertial modes by Lockitch \& Friedman (1999).
All of these calculations were carried out by means of an eigenmode
analysis, i.e. by assuming an oscillatory constant-frequency solution.
In contrast, in this paper we will perform \emph{time evolutions} of
the linearised equations of motion of a rigidly rotating Newtonian
polytrope to investigate oscillatory behaviour in the time domain.
This will allow us to check, as accurately as a time evolution
approach permits, the f-mode, r-mode and inertial-mode frequencies and
eigenfunctions in the literature, and also to extend the calculation
of the inertial mode frequencies into the regime of rapid rotation.
Our work is in much the same spirit as that of Papaloizou \& Pringle
(1980), who performed time evolutions of r-modes in rapidly rotating
thin shells.

Our long-term aim is to provide a stable and well-tested platform for
developing a more advanced time evolution code, capable of evolving more
realistic stellar models.  For instance, future modifications will include
evolving a star with a differentially rotating background.  This issue is
of particular interest, as recently there has been speculation that no mode
resembling an r-mode will exist in a sufficiently strongly differentially
rotating star (Karino et al.\ 2000).  By starting with a rigidly rotating
configuration, and tracking the frequency of a r-mode through a series of
runs where the degree of differential rotation is increased step-by-step, it
should be possible to probe the mode structure in highly differentially
rotating stars.

Another problem that could be addressed using our code is that of modelling
gravitational radiation reaction as a local force, included in the
equations of motion.  A number of elegant formulations of this problem have
been presented (see Blanchet et al.\ 1990 and Rezzolla et al.\ 1999), but
have yet to be fully applied to the modes of compact stars.  Recently,
Lindblom, Tohline \& Vallisneri (2001) included radiation reaction in their
evolutions of the r-modes using formulae from Rezzolla et al.\, but made
some approximations when calculating the force.  With our computationally
less intensive linear code one can attempt to compare the results
obtained via their approximate treatment with those obtained from the full
formalism, before the mode saturates due to nonlinear effects..

Another future modification of our code could be to include the first
post-Newtonian corrections to the equilibrium star and the equations of
motion.  This problem is also of considerable interest, as recently Ruoff
\& Kokkotas (2001) have suggested that r-modes will not exist in
sufficiently compact relativistic neutron stars.  They speculate that this
is due to frame dragging, an effect which is present at first
post-Newtonian order.  This hypothesis could be tested using a strategy
similar to that described above for differential rotation; specifically by
performing a series of evolutions of successively more and more compact
stars, tracking the r-mode between each evolution.

The implementation of all of these pieces of physics are challenging,
and we hope to learn how to deal with each separately in the linear
regime before combining them or extending them to the non-linear
regime.  In essence, our strategy is a modular one, with each problem
being isolated and studied on its own.  (As regards non-linear
evolutions, by coupling several linear time evolution codes it will
be possible to monitor mode-mode coupling in the weakly non-linear
regime).

In this paper we postpone such advanced calculations, and present
evolutions of rigidly rotating Newtonian polytropes, with the
perturbations in the gravitational potential neglected (the Cowling
approximation).  Our goal is to demonstrate the accuracy and stability
of our code prior to studying the more complicated physical problems
discussed above.  We begin by describing some of the computational
details (section \ref{sect:fotp}) before examining the f-modes (section
\ref{sect:tfm}), the r-modes (section \ref{sect:trm}) and finally
inertial modes (section \ref{sect:grm}).

\section{Formulation of the problem}
\label{sect:fotp}

\subsection{The unperturbed star}
\label{sect:tus}

The unperturbed rotating stellar structure was calculated using the Hachisu
self-consistent field method (Hachisu 1986).  The equation of state is
given by the usual polytropic form:
\begin{equation}
P = K \rho^{\gamma},
\end{equation}
where $P$ and $\rho$ denote pressure and density, while $K$ and $\gamma$
are constants.  The polytropic exponent $\gamma$ is related to the polytropic
index $n$ by 
\begin{equation}
\gamma = 1 + 1/n.
\end{equation}
Each equilibrium model is characterised by its central density and
ratio of the polar to equatorial axes. In constructing an equilibrium
model, the equations governing the gravitational potential and the
hydrostatic equilibrium are solved iteratively in an integral form
(starting from a non-rotating model) until convergence to a desired
numerical accuracy is achieved. Our code reproduces the numerical
models in Hachisu (1986) to the number of significant digits
published.

We will find it useful to introduce a coordinate system consisting of the
usual spherical polar $\theta$ and $\phi$ coordinates, but with a modified
radial coordinate $x=x(r,\theta)$ which is fitted to surfaces of constant
pressure of the background star.  With this choice, the unperturbed
pressure and density are functions of $x$ only, i.e. $P = P(x)$ and $\rho =
\rho(x)$.  The usual spherical polar partial derivatives can then be
written as:
\begin{eqnarray}
\label{s2}
\left.\frac{\partial}{\partial r}\right|_\theta &=&
\left.\frac{\partial x}{\partial r}\right|_\theta
\left.\frac{\partial}{\partial x}\right|_\theta, \\
\label{s3}
\left.\frac{\partial}{\partial \theta}\right|_r &=&
\left.\frac{\partial x}{\partial \theta}\right|_r
\left.\frac{\partial}{\partial x}\right|_\theta +
\left.\frac{\partial}{\partial \theta}\right|_x.
\end{eqnarray}
We will choose to set $x$ equal to $r$ in the equatorial plane,
i.e. $x(r,\pi/2)=r$.  For $\theta < \pi/2$, the oblate shape of the star
will mean that $x(r,\theta) > r$.  The partial derivatives of $x$ with
respect to $r$ and $\theta$ can be easily evaluated using:
\begin{equation}
\label{dxbdr}
\left.\frac{\partial x}{\partial r}\right|_\theta = 
\left.\frac{\partial \rho}{\partial r}\right|_\theta 
\left/ \frac{\partial \rho}{\partial r}\right|_{\theta = \pi/2},
\end{equation}
\begin{equation}
\label{dxbdt}
\left.\frac{\partial x}{\partial \theta}\right|_r = 
\left.\frac{\partial \rho}{\partial \theta}\right|_r 
\left/ \frac{\partial \rho}{\partial r}\right|_{\theta = \pi/2}.
\end{equation}
This choice of coordinate system will make it particularly easy to
implement the outer boundary conditions described in section
(\ref{sect:bc}).

\subsection{Equations to be solved}

For a given background star, three pieces of physics are required to
describe this problem: The equations of motion, the equation of mass
conservation, and an equation of state for the perturbations.

The equations of motion referred to the frame of the rotating background are:
\[
\frac{\partial {\bf v}}{\partial t} + ({\bf v} \cdot \nabla){\bf v} 
+ 2 {\bf \Omega} \times {\bf v} 
+ {\bf \Omega}\times({\bf \Omega \times r})
\]
\begin{equation}
\label{exacteom}
\hspace{5cm} 
= -\frac{1}{\rho} \nabla P - \nabla \phi,
\end{equation}
where $\bf v$ denotes the departure of the fluid from the rigid
rotation of the background equilibrium model.  Performing an Eulerian
perturbation and making the Cowling approximation leads to:
\begin{equation}
\label{lineom}
\frac{\partial {\bf v}}{\partial t} + 2 {\bf \Omega} \times {\bf v}
= \frac{\delta \rho}{\rho^2} \nabla P - \frac{1}{\rho} \nabla \delta P.
\end{equation}
The prefix $\delta$ denotes an Eulerian perturbation, while $\rho$ and $P$
refer to the background star.  The equation of continuity is:
\begin{equation}
\frac{\partial \rho}{\partial t} + \nabla \cdot (\rho {\bf v}) = 0
\end{equation}
which linearises to give:
\begin{equation}
\label{lincont}
\frac{\partial \delta \rho}{\partial t} + \nabla \cdot (\rho {\bf v}) = 0.
\end{equation}
We will write the equation of state for the perturbations as:
\begin{equation}
\label{lineos}
\frac{\Delta P}{P} = \gamma_{\rm pert}  \frac{\Delta \rho}{\rho},
\end{equation}
where $\Delta$ denotes a Lagrangian perturbation.  Using the relation that,
for a displacement vector ${\bf \xi}$, the Lagrangian and Eulerian
perturbations of a scalar are related by
\begin{equation}
\Delta = \delta + \bxi \cdot \nabla
\end{equation}
we find:
\begin{equation}
\frac{\delta \rho}{\rho} = \frac{1}{\gamma_{\rm pert}} \frac{\delta P}{P}
                          -\bxi \cdot {\bf A}.
\end{equation}
The  vector $\bf A$ is given by:
\begin{equation}
\label{s1}
{\bf A} = \nabla \log \rho - \frac{1}{\gamma_{\rm pert}} \nabla \log P,
\end{equation}
and its magnitude is known as the Schwarzschild discriminant, which depends
upon the background pressure and density, and on the index $\gamma_{\rm
pert}$ describing the perturbations.  We will consider the case where the
perturbations obey the same equation of state as the background, so that
$\gamma_{\rm pert} = \gamma$ and
\begin{equation}
{\bf A} = 0.
\end{equation}
As discussed in the stellar oscillation literature (see e.g. Tassoul [1978]
or Unno et al.\ [1989]), the magnitude and direction of ${\bf A}$ determines
the nature of g-modes in the star.  If ${\bf A}$ points inwards the star has
stable oscillatory g-modes, while if ${\bf A}$ points outwards the g-modes
are unstable, and the star is described as being `convectively unstable'.
In choosing ${\bf A = 0}$ there exists no restoring force at all for g-mode
type displacements, so we have eliminated the g-modes altogether from the
stellar spectrum.

Expressed in terms of  $\delta P$, equations
(\ref{lineom}) and (\ref{lincont}) become:
\begin{eqnarray}
\label{lineoma}
\frac{\partial {\bf v}}{\partial t} + 2 {\bf \Omega} \times {\bf v}
&=& \frac{\nabla P}{\rho} 
\frac{1}{\gamma} \frac{\delta P}{P} 
- \frac{1}{\rho} \nabla \delta P, \\
\label{linconta}
\frac{\partial \delta P}{\partial t} &=& - \frac{\gamma P}{\rho} \nabla
\cdot (\rho {\bf v}).
\end{eqnarray}
With respect to the standard spherical polar vector basis $[\bf{e_r,
e_\theta, e_\phi}]$ (\emph{not} the coordinate vector basis) we then find:
\begin{eqnarray}
\label{id1}
\frac{\partial f_r}{\partial t} &=& -\frac{\partial \delta P}{\partial r} 
+2 \Omega \sin \theta f_\phi + \frac{\partial P}{\partial r}
\frac{1}{\gamma}\frac{\delta P}{P}, \\
\label{id2}
\frac{\partial f_\theta}{\partial t} &=& 
-\frac{1}{r}\frac{\partial \delta P}{\partial \theta}
+2 \Omega \cos \theta f_\phi 
+ \frac{1}{r}\frac{\partial P}{\partial \theta}
\frac{1}{\gamma}\frac{\delta P}{P}, \\
\label{id3}
\frac{\partial f_\phi}{\partial t} &=& 
-\frac{1}{r\sin \theta}\frac{\partial \delta P}{\partial \phi}
-2 \Omega(\cos \theta f_\theta + \sin \theta f_r), \\
\label{id4}
\frac{\partial \delta P}{\partial t} &=& -\frac{\gamma P}{\rho}
\nabla \cdot {\bf f}.
\end{eqnarray}
where the velocity $\bf v$ has been replaced by the flux ${\bf f}= \rho \bf
v$, as this choice of variable leads to very simple outer boundary
conditions (see section \ref{sect:bc}).

We follow Papaloizou \& Pringle (1980) 
in decomposing any given perturbation as a
sum over basis functions of the form $\cos m \phi$ and $\sin m \phi$, where
$m$ is an integer, e.g. for the pressure:
\[
\delta P(t, r, \theta, \phi) = \sum_{m=0}^{m=+\infty} 
                               \delta P^{+}_m(t, r, \theta) \cos{m \phi}
\]
\begin{equation}
\label{eq:phidec}
\hspace{4cm}    +  \delta P^{-}_m(t, r, \theta) \sin{m \phi}.
\end{equation}
This allows us to reduce the number of space dimensions that appear in the
numerical evolution from three to two.  With the understanding that the
perturbation variables are now functions of $(t,r,\theta)$, but not $\phi$,
equations (\ref{id1})--(\ref{id4}) then apply to each azimuthal number $m$
separately, i.e. the equations decouple in $m$, giving a total of eight
equations.  The partial derivatives with respect to $\phi$ are trivial to
calculate analytically.  In summary, equations (\ref{id1})--(\ref{id4})
then give a set of eight coupled first order linear partial differential
equations in the eight unknowns $(\delta P^+, \delta P^-, f_r^+, f_r^-,
f_\theta^+, f_\theta^-, f_\phi^+, f_\phi^-)$.

For a perturbation of definite azimuthal number $m$, this decomposition
with respect to $\phi$ greatly reduces the memory storage and running time
of the code, as compared to a fully three-dimensional computation.  The
fundamental reason that the equations decoupled is that the background star
is axisymmetric.  A further decomposition in which the $\theta$ dependence
of the perturbation variables is written in terms of a set of spherical
harmonic functions $Y_{lm}$ is not very useful---the non-spherical nature
of the rotating unperturbed star would couple terms with different $l$
indices, with the coupling becoming stronger the more rapidly the star
rotates.  By eliminating only the $\phi$ dependence, we have simplified the
computational costs of the problem to the maximum extend possible without
significantly increasing the complexity of the equations to be solved.

\subsection{Boundary conditions}
\label{sect:bc}

By definition, the Lagrangian perturbation in the pressure is zero at the
surface, i.e.
\[
\delta P + \bxi \cdot \nabla P = 0 {\rm \hspace{1cm} (at \, surface).}
\]
For the unperturbed star, equation (\ref{exacteom}) gives:
\begin{equation}
\label{unpert}
\nabla P = -\rho{\bf \Omega} \times ({\bf \Omega \times r})
           -\rho \nabla \phi.
\end{equation}
For a polytrope $\rho = 0$ at the surface, from which it follows that
$\nabla P=0$ at the surface.  It follows at once that our outer boundary
condition is:
\begin{equation}
\delta P = 0 {\rm \hspace{1cm} (at \, surface).}
\end{equation}
Our choice of coordinates makes imposing the outer boundary condition very
simple: We have $\delta P(x=R) = 0$, where $R$ denotes the equatorial
radius.  Also, the mass flux ${\bf f} = \rho {\bf v}$ is zero at the
surface, by virtue of the vanishing density, and so
\begin{equation}
{\bf f} = {\bf 0}  {\rm \hspace{1cm} (at \, surface).}
\end{equation}
Our choice of surface fitted coordinates and perturbation variables
guarantees that we never need to evolve quantities at the surface of the
star.  Numerically, this is highly desirable, as resolving the possibly
rapidly falling density profile near the surface can lead to
inaccuracies and instabilities.

As we will consider perturbations with $m>0$ we can set $\bf f =0$ at the
centre.

\section{Results}
\label{sect:nd}

\begin{figure}
   \centerline{ \psfig{file=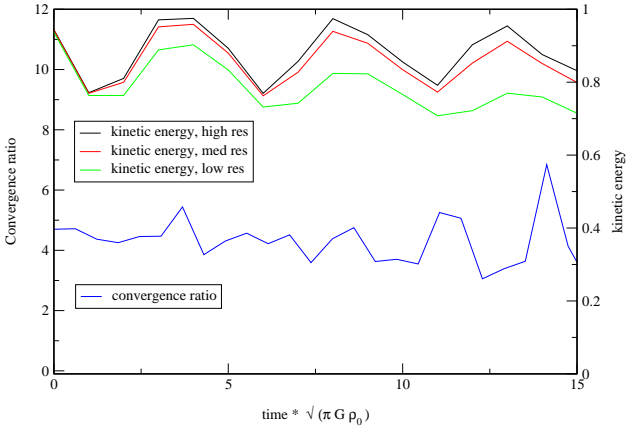,width=9cm,angle=-90} }
   \caption{Demonstration of second order convergence.  The three upper
   curves plot the kinetic energy at three different grid resolutions,
   labelled as high, medium and low resolutions.  The lower curve is the
   dimensionless convergence ratio, defined in equation (\ref{eq:conv}).
   Time is plotted in unity of $1/\sqrt{\pi G \rho_0}$, where $\rho_0$ is
   the average density of the non-rotating star of the same mass.}  
\label{fig:conv3}
\end{figure}

Equations (\ref{id1})--(\ref{id4}), decomposed with respect to $\phi$ as
described above (equation [\ref{eq:phidec}]) were propagated forward in
time using the two-step Lax-Wendroff scheme described in Press et al.\
(1986).  The code is second order convergent, and runs stably for hundreds
of rotation periods.  The second order convergence is illustrated by figure
\ref{fig:conv3}.  In this figure we plot the total kinetic energy of the
star, as measured in the rotating frame, as a function of time for some
arbitrary initial data.  The time is given in units of $1/\sqrt{\pi G
\rho_0}$, where $\rho_0$ is the average density of the non-rotating star of
the same mass and equation of state.  The star rotates at a rate $\Omega /
\sqrt{\pi G \rho_0} = 0.538$, corresponding to a polar-to-equatorial axis
ratio of $0.79$.  The evolution was performed for three different grid
resolutions, labelled as high, medium and low in the figure, with the
resolution differing by a factor of two between each successive run.  In
full, for the high resolution run we used a grid of $64$ angular points by
$60$ radial ones, the medium resolution grid was $32$ by $30$ points, and
the low resolution grid was $16$ by $15$ points.  The kinetic energy is
normalised to unity at $t=0$.  A small oscillation is visible,
corresponding to the excitation of a p-mode (Cox 1980), whose exact nature
need not concern us here.  A convergence ratio $C$ is then calculated as
the ratio:
\begin{equation}
\label{eq:conv}
C  = \frac
{E(t, {\rm high \, resolution}) - E(t, {\rm low \, resolution})}
{E(t, {\rm high \, resolution}) - E(t, {\rm medium \, resolution})}
\end{equation}
For a code with second order accuracy in space, the ratio can be shown to
be approximately equal to 5.  This is indeed the case for our time
evolution code.  Beyond the time interval shown in the figure, 
the oscillations in the energy start to loose phase
coherence, and the convergence ratio begins to oscillate strongly.  The
convergence ratio then looses its significance.

\begin{figure}
   \centerline{ \psfig{file=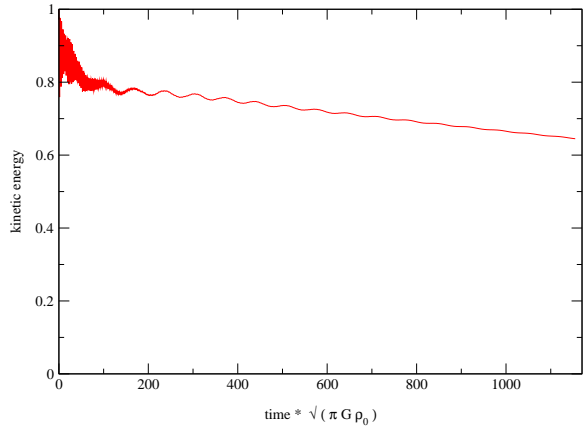,width=9cm,angle=-90} }
   \caption{Kinetic energy as a function of time, normalised to unity at
   $t=0$, and time measured in the same units as in figure
   \ref{fig:conv3}).  Around $100$ stellar rotations occur over the period
   plotted.}
\label{fig:ke_stability}
\end{figure}

The long-term stability of the code is illustrated by figure
\ref{fig:ke_stability}, which plots the kinetic energy of the star as a
function of time.  The rotation rate is the same as above, and the star
rotates $100$ times over the period plotted.  A small amount of artificial
viscosity was included to give this long-term stability, and the grid
consisted of $32$ angular points and $30$ radial ones.  The initial data
used was that of an $l=m=2$ r-mode for a slowly rotating star (about $80$
r-mode oscillations occur over this interval).  A rather sharp loss of
kinetic energy occurs over the first ten or so rotations.  This is likely
due to the initial data exciting a number of short-wavelength modes whose
energy was rapidly dissipated.  Over the remainder of the evolution, the
kinetic energy remains approximately constant, falling by about $10\%$ over
the remaining $90$ rotations, confirming the long-term stability of our
code.  This evolution required approximately two hours of computer time on
a standard PC.  Most of the results in this paper are based on runs of this
duration or less.

Part of the above energy loss is due to the artificial viscosity, and part
due to numerical error.  Given that the code has been shown to be second
order convergent, an increase in grid resolution by a factor of $N$ would
reduce the rate of energy loss due to numerical error by a factor $N^2$.
On the basis of runs we have performed where the artifical viscosity was
set to zero, we estimate that an increase in grid resolution by a factor of
$10$ (i.e. a $320$ by $300$ grid) would be sufficient to reduce the
energy loss to around $1\%$ for an evolution of the above duration.  Such a
run would take over a week on a standard PC, and so has not been performed
here.  However, should such a high level of accuracy be required for a
particular application, the necessary runs could be easily be done on a
supercomputer.

In the following sections we will use this time evolution code to
study f-modes, r-modes and inertial modes.  We will use as initial
data some (possible crude) approximation to the true eigenfunction.
The perturbed quantities will then be Fourier analysed.  By performing
a series of such evolutions, of gradually increasing stellar rotation
rate, the mode of interest can then be tracked from the slow-rotation
case (where it's approximate frequency is already known) into the
rapid rotation regime.  

Note that for an evolution of duration $T$, the error in the frequency
localisation of the transform is approximately equal to $1/T$.  It was
therefore necessary to perform long time evolutions to obtain accurate
frequencies.  This error in measurement is indicated by
error bars on the mode frequency plots that follow (figures
\ref{fig:lm2}--\ref{fig:inertial}).  However, it should be noted that
this is not the only source of error---the finite grid resolution will
affect the results also, but its effect is not as easy to quantify.
The error bars in the figures should be regarded as simple estimates of the
error.

Note that all mode frequencies given in this paper are measured in the
rotating frame.  Also, we only present results for the case when the
polytropic index $n$ is equal to unity.

\subsection{The f-modes}
\label{sect:tfm}

As described above, to investigate a particular mode it was necessary
to supply our code with initial data that would excite it.  In the
case of f-modes we chose to use initial data of the form:
\begin{equation}
\delta P = \rho \left(\frac{r}{R(\theta)}\right)^l Y_{ll},
\end{equation} 
where $R(\theta)$ denotes the stellar radius at polar angle $\theta$
and $Y_{ll}$ is a spherical harmonic.  This reduces to the correct
f-mode pressure perturbation in the limit of zero rotation for an
incompressible star, in the Cowling approximation.  The perturbed mass
flux $\bf f$ was set to zero in the initial data.  The mode frequency
was then extracted by taking power spectra of the time series of the
perturbed quantities. A more efficient excitation of the f-modes would
have been possible if non-zero initial data was provided for $\bf f$
also, but this was not found to be necessary---the f-modes were easily
identifiable using the above initial data.  Indeed, this scheme
efficiently excited the $l=m$ mode in all but the most rapidly
rotating stars.  

\begin{figure}
   \centerline{ \psfig{file=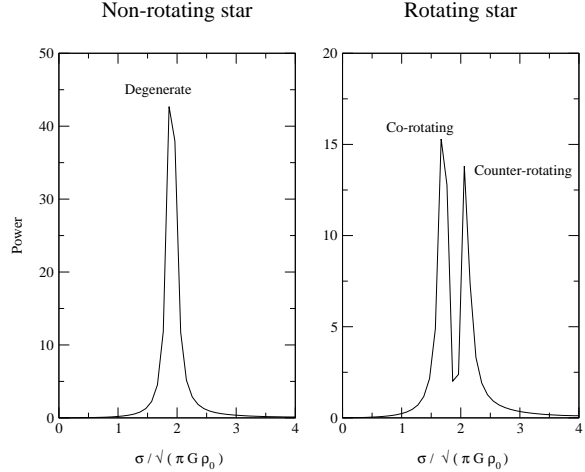,width=9cm,angle=-90} } 
   \caption{Power
   spectra of the perturbed pressure $\delta P^+$ for a non-rotating star
   (left panel) and for a rotating star (right panel).  The powers
   are in the same (arbitrary) units in both panels, while the mode
   frequencies are in units of $\sqrt{\pi G \rho_0}$, as indicated.}
   \label{fig:fmodefft}
\end{figure}

The power spectra of the evolutions consisted of just one peak
for the non-rotating star.  This then split into two peaks for $\Omega \not
= 0$, corresponding to the breaking of the degeneracy between co- and
counter-rotating modes.  This is illustrated in figure \ref{fig:fmodefft}
for the spectra of $\delta P^+$ for the $l=m=2$ f-mode.  The left hand
panel was obtained for a non-rotating star, while the right hand one was
for a rotating one ($\Omega / \sqrt{\pi G \rho_0} = 0.37$,
polar-to-equatorial radius ratio 0.92).  Note how the power in the
degenerate $\Omega = 0$ peak for the static star is split roughly equally
between the co- and counter-rotating peaks in the $\Omega \not = 0$ case.
As the star's rotation rate was increased, the frequency separation of the
modes increased, and the power was shared less evenly between the two.

Before examining the case of rotating stars in detail we will consider the
simpler case of non-rotating ones.  The mode frequencies, as calculated by
a normal mode analysis in the Cowling approximation (Kokkotas, personal
communication), are given in table \ref{table:fmode} for the $l=m=2$, $4$
and $6$ modes.  Also shown are the mode frequencies we obtained from our
time evolution code, and the percentage difference between the two.  In
this non-rotating case the eigenmode calculation is known to be highly
accurate, so that the percentage differences in the table give an
indication of the accuracy of our code.  The difference is very small (less
than $1\%$) in the $l=m=2$ case, but somewhat higher (around $4\%$) for the
high-index modes.  This decrease in accuracy is almost certainly due the
higher-index modes varying on smaller length scales than the $l=m=2$ mode,
and therefore being less well represented on our spatial grid.

\begin{center}
\begin{table}
%\footnotesize
\label{table:fmode}
\begin{tabular}{llll} \hline
$l=m=$ & $2$ & $4$ & $6$ \\
Eigenmode             & $1.912$ & $2.577$ & $3.077$ \\
Time evolution        & $1.90$  & $2.68$  & $2.96$   \\
Percentage difference & $0.5$   & $3.8$   & $3.9$ \\
\hline
\end{tabular}
\caption{F-mode frequencies as calculated by an eigenmode analysis, and
using our time evolution code (units same as in figure \ref{fig:fmodefft}).
The eigenmode vales are anticipated to be of high precision, so that the
percentage difference between the two is an indication of the accuracy of
our time evolutions. }
\end{table}
\end{center}

We then investigated f-modes in rotating stars.  The f-mode
eigenfrequencies and eigenfunctions of rapidly rotating stars have
been calculated by Ipser \& Lindblom (1990), without making the
Cowling approximation.  Specifically, they considered polytropic
stars, and searched for modes which had spherical harmonic indices
$l=m$ in the limit of slow rotation.

\begin{figure}
   \centerline{ \psfig{file=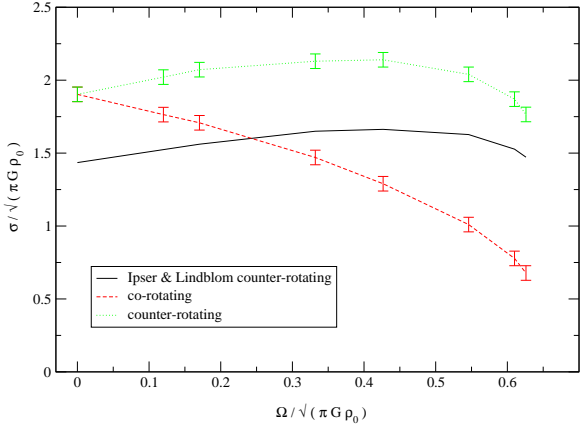,width=9cm,angle=-90} } \caption{The
   $l=m=2$ f-mode frequencies as a function of rotation rate.  The upper
   curve is for the counter-rotating mode as calculated by our time
   evolution code.  This is to be compared with the Ipser \& Lindblom
   result of the solid curve.  The difference between the two results is
   due to our having made the Cowling approximation.  The monotonically
   decreasing curve is the co-rotating mode frequency as calculated by our
   code.  All frequencies are in units of $\sqrt{\pi G \rho_0}$, where
   $\rho_0$ is the average density of the non-rotating star of the same
   mass and equation of state.  In these units, the Kepler rotation rate is
   2/3.}  \label{fig:lm2}
\end{figure}

\begin{figure}
   \centerline{ 
   \psfig{file=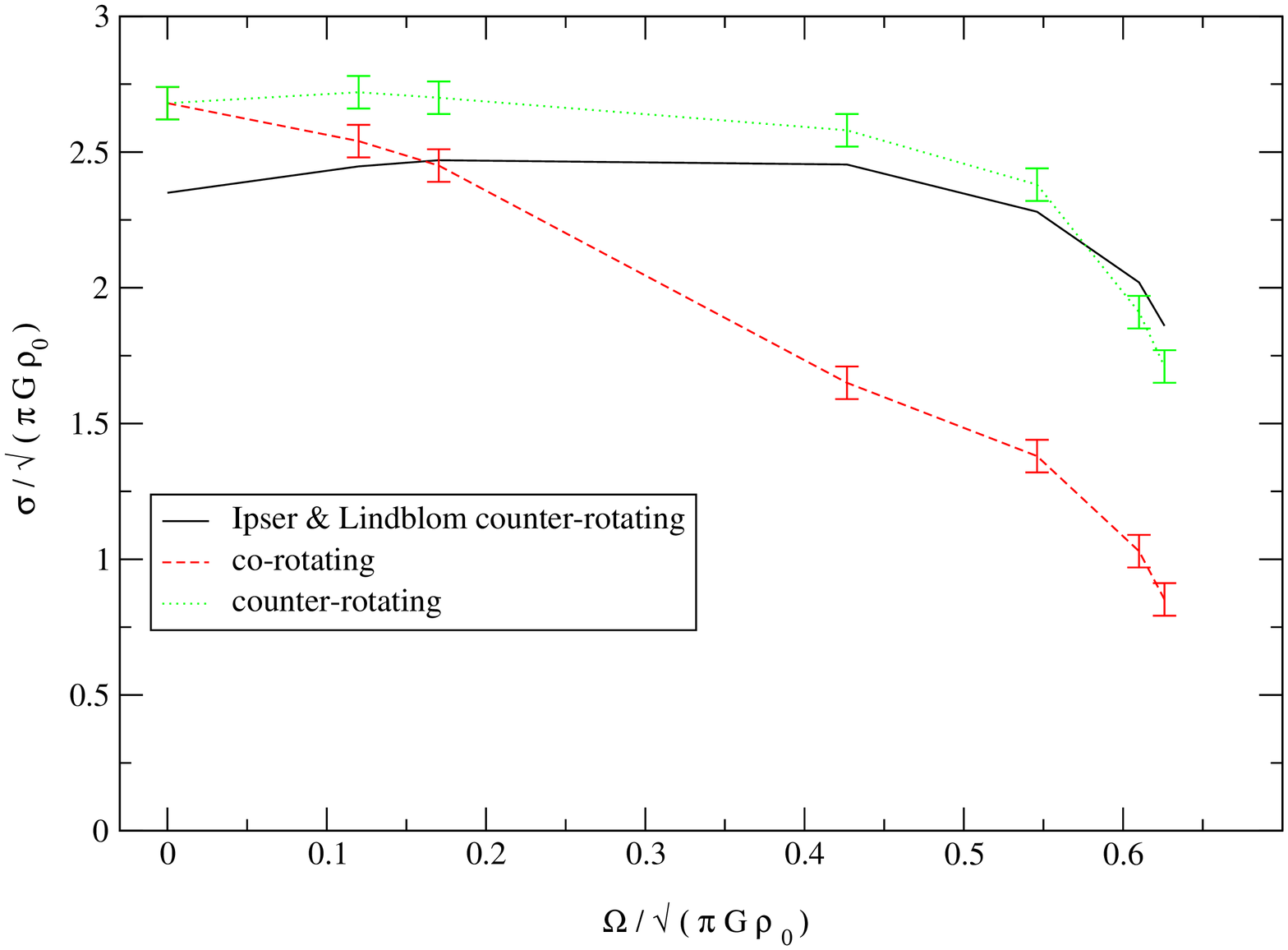,width=9cm,angle=-90} } 
   \caption{As figure \ref{fig:lm2}, with $l=m=4$}  
   \label{fig:lm4}
\end{figure}

\begin{figure}
   \centerline{ 
   \psfig{file=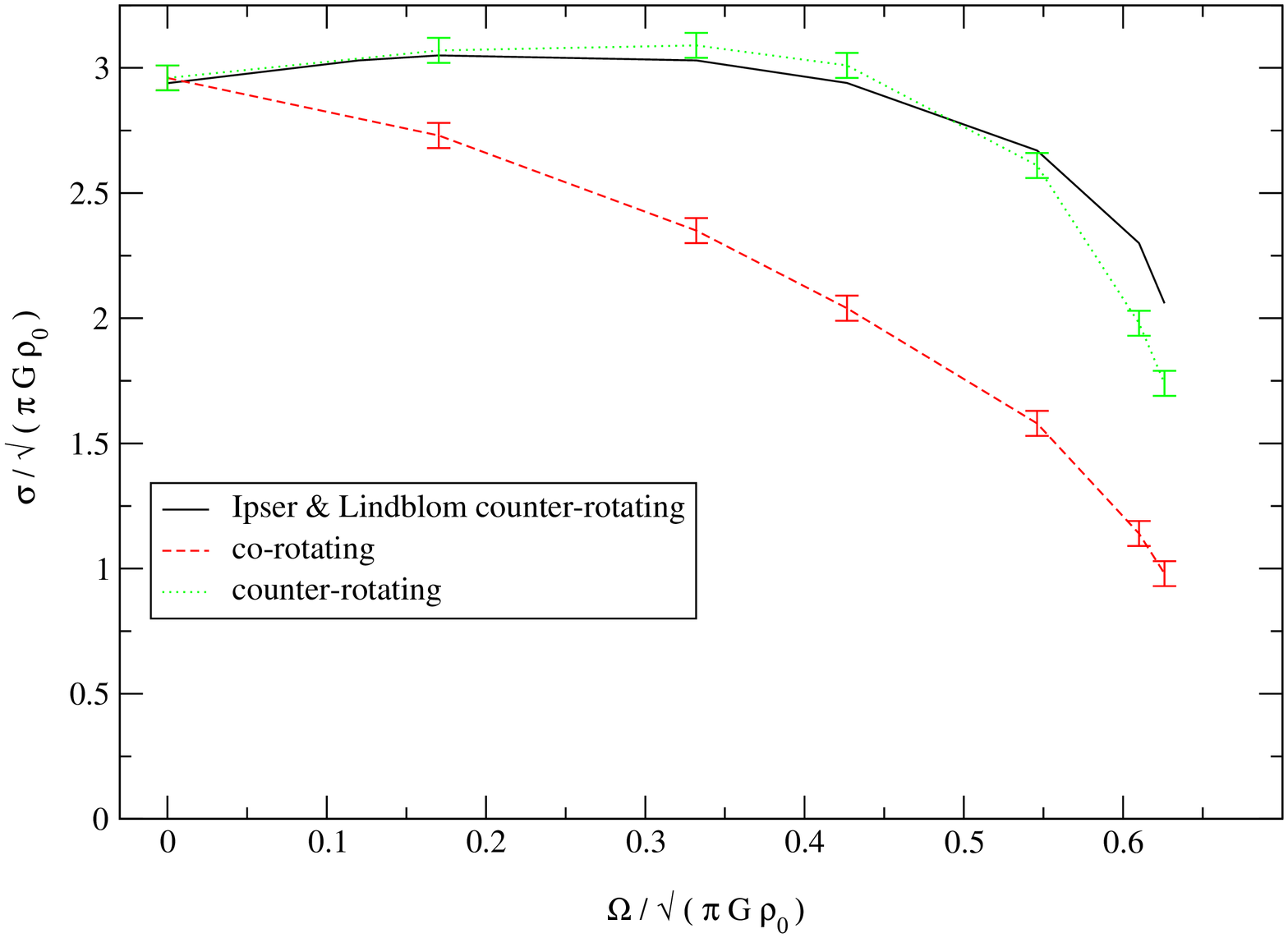,width=9cm,angle=-90} } 
   \caption{As figure \ref{fig:lm2}, with $l=m=6$}  
   \label{fig:lm6}
\end{figure}

The frequency of the co- and counter-rotating modes, together with the
Ipser \& Lindblom values for the counter-rotating mode, are plotted in
figures \ref{fig:lm2}, \ref{fig:lm4} and \ref{fig:lm6}, for $l=m=2,4,6$
respectively.


For most rotation rates the counter-rotating mode frequencies
calculated from our time evolutions are somewhat higher than those
calculated by Ipser \& Lindblom.  This is almost certainly due to our
having made the Cowling approximation, while Ipser \& Lindblom have
not.  The discrepancy is typically of order 30\% for $l=m=2$, 10\% for
$l=m=4$ and 3\% for $l=m=6$.  Such a decrease of discrepancy with
increasing harmonic index is to be expected: higher order modes have
alternating regions of positive and negative density perturbation,
whose contribution to the perturbation in the gravitational potential
at a given point tend to cancel one-another out (Cowling 1941).  We
can therefore conclude that, within the error margin of the Cowling
approximation, our time evolution code confirms the f-mode frequencies
calculated by Ipser \& Lindblom.

\subsection{The r-modes}
\label{sect:trm}

Because of the considerable interest aroused by the r-modes (see Andersson
\& Kokkotas 2001 for a recent review), we used our time evolution code to
examine both the frequencies and eigenfunctions of this class of mode.  For
initial data we provided a perturbation in the mass flux of the form:
\begin{equation}
\label{eq:rmid}
{\bf f} = \rho \left(\frac{r}{R(\theta)}\right)^l {\bf Y^B}_{ll},
\end{equation}
where ${\bf Y^B}_{ll}$ is a magnetic spherical harmonic (Thorne 1980).  In
the limit of slowly rotating stars, this becomes an exact mode solution of
the linearised equations of motion.

\subsubsection{The r-mode frequencies}

\begin{figure}
   \centerline{ \psfig{file=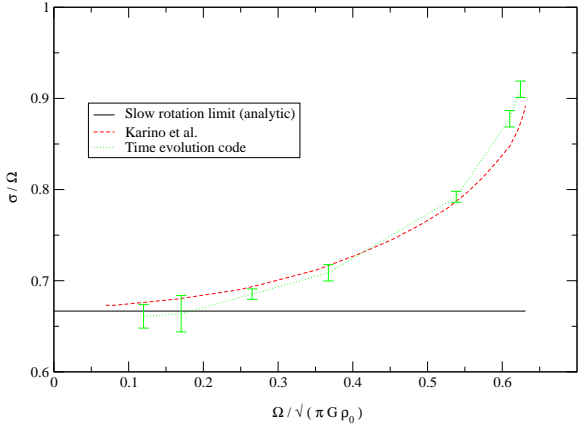,width=9cm,angle=-90} } \caption{The
   frequency of the $l=m=2$ r-mode as a function of rotation rate.  Mode
   frequencies are given in units of $\Omega$, the stellar rotation rate.
   As in previous plots, the error bars indicated the result of our time
   evolution code.  The dotted line gives the mode frequency calculated by
   Karino et al.\ (2000). The horizontal line indicates the analytic slow
   rotation limit of 2/3.  }  \label{fig_sigma}
\end{figure}

In order to extract r-mode frequencies, the power spectra of the
$\theta$ and $\phi$ components of the mass flux were analysed.  The
result of a series of such evolutions for the $l=m=2$ r=mode is shown
in figure \ref{fig_sigma}, with the error bars indicating the
uncertainty in the frequency.  On the vertical axis we plot the
dimensionless ratio $\sigma/\Omega$, where $\sigma$ is the mode
frequency.  On the horizontal axis we plot the normalised rotation
rate $\Omega/\sqrt{\pi G \rho_0}$.
For comparison, a curve showing the mode frequency as calculated by
Karino et al.\ (2000) is shown also, together with a horizontal line
at the mode frequency as calculated using the slow rotation limit
($\sigma/\Omega = 2/3)$.  Note that Karino et al.\ did \emph{not}
make the Cowling approximation.  Within the indicated uncertainties
due to the finite duration of the time evolution, the data sets agree
very well, even at the fastest rotation rates.  To leading order in
the rotation rate, the Cowling approximation is known to not affect
the mode frequency, so this close agreement was to be expected.  There
is a slight difference between our results and those of Karino et al.\
at the highest rotation rates.  We suspect this is due to the Cowling
approximation being a poorer approximation in this regime, as the
fractional density perturbation $\delta \rho /\rho$ is known to
increase with rotation rate as $\Omega^2$ for r-modes (Andersson \&
Kokkotas 2001).

\subsubsection{The r-mode eigenfunctions}

The r-mode eigenfunctions were calculated by Karino et al.\ (2000).  In order
to confirm their veracity, these eigenfunctions were used as initial data
for our time evolutions.  An accurate eigenfunction would excite the
appropriate r-mode only, with little excitation of other modes, while an
inaccurate function would deposit a significant quantity of energy over
other modes.  In order to quantify this, we examined the power spectra of
several evolutions with different stellar rotation rates, for the $l=m=2$
r-mode.  For each evolution we noted the ratio of the r-mode excitation
peak to the next strongest excitation.  

We found that in the power spectra of $f_\theta$ and $f_\phi$, the
r-mode excitation was extremely clean, so that secondary peaks are
many orders of magnitude smaller.  For $f_r$ and $\delta p$
significant secondary peaks were present, implying that our initial
data also excited some radial modes or polar modes of oscillation.
The ratios of these to the r-mode peak are given in table
\ref{table_peaks}, for stars with rotation rates $\Omega / \sqrt{\pi G
\rho_0}$ of $0.170$, $0.538$ and $0.624$.
\begin{center}
\begin{table}
%\footnotesize
\label{table_peaks}
\begin{tabular}{llll} \hline

$\sigma / \sqrt{\pi G \rho_0}=$ & $0.170$ & $0.538$ & $0.624$ \\

& \multicolumn{3}{c}{Peak ratio}  \\
         
\hline   
$f_r^+$ &  224 & 42.6 & 4.26 \\
$ \delta P^+$ &  7740 & 174 & 8.06 \\
\hline
\end{tabular}
\caption{Ratio of $m=2$ r-mode excitation peak to next strongest peak
(representing other modes), for
the quantities $f_r^+$ and $\delta P^+$, for three different rotation rates.
Recall than the Kepler rotation rate is $2/3$ in these units.}
\end{table}
\end{center}

Clearly, in all cases the initial data efficiently picks out the r-mode
from the stellar spectrum.  The excitation is more accurate the slower the
rotation of the star, but even in the the rapidly rotating $\Omega /
\sqrt{\pi G \rho_0} = 0.624$ case there is little other mode excitation.
The kinetic energy of the perturbations is quadratic in the mass flux, so
that in this case only a few percent of the kinetic energy of the initial
data is deposited into modes other than the r-mode.  For the relatively
slowly rotating $\Omega / \sqrt{\pi G \rho_0} = 0.170$ case, only a few
thousandths of a percent is deposited.  We can therefore confirm that the
eigenfunctions computed by Karino et al.\ (2000) are accurate.

\subsection{Inertial modes}
\label{sect:grm}

The r-modes of the previous section are a special case of a wider class of
oscillations, known as \emph{inertial modes} (Greenspan 1968, Pedlosky 1987). 
 The defining feature of the
inertial modes is that their frequency vanishes linearly in the limit of
zero stellar rotation, or equivalently that $\sigma / \Omega$ remains
finite in this limit.  The r-modes of the last section are the subset of
the inertial modes which have purely axial velocity perturbations in the
limit of zero rotation (axial means that the velocity can be written in
terms of $\nabla \times Y_{lm}$).  Furthermore, for stars with zero
Schwarzchild discriminant, the two spherical harmonic indices of the
r-modes must be equal, i.e. $l=m$ (Provost et al.\ 1981).  

In contrast, inertial modes are hybrid, having velocity perturbations with
both axial and polar parts, even in the slow rotation limit (polar means
that the velocity field can be written as a sum of terms proportional to
$ Y_{lm} {\bf e}_r$ and $\nabla Y_{lm}$).  Also, the inertial modes are not
restricted to $l=m$ spherical harmonics.  Inertial modes were considered in
detail recently by Lockitch \& Friedman (1999), who calculated mode
frequencies and eigenfunctions.  It was found that many of these inertial
modes could undergo the CFS instability and are therefore interesting from
a gravitational wave point of view, even though they are probably not as
strongly unstable as the $l=m=2$ r-mode.  It is therefore of interest to
extend the calculations of Lockitch \& Friedman and calculate these mode
frequencies for more rapidly rotating stars.

We have performed such an analysis using our time evolution code.
We used the analytic approximations to eigenfunctions given by Lockitch \&
Friedman as initial data to carry out evolutions of several inertial modes.

\begin{figure}
   \centerline{ \psfig{file=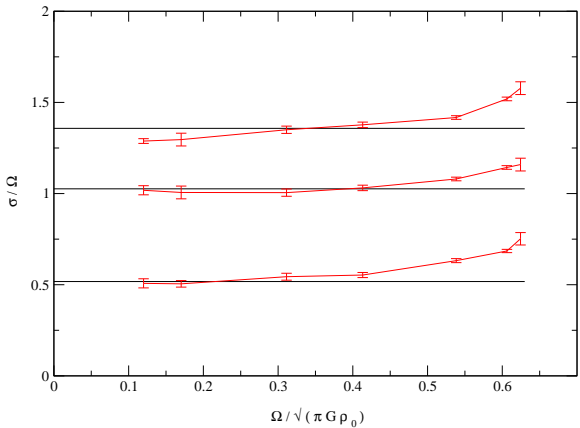,width=9cm,angle=-90} }
   \caption{Oscillation frequencies for inertial modes.  The three modes
   all have leading order velocity perturbations proportional to the $l=4$,
   $m=2$ spherical harmonic in the limit of slow rotation, but have
   different radial behaviour.  The units are the same as in figure
   \ref{fig_sigma}.  The horizontal lines indicate the mode frequencies as
   calculated by Lockitch \& Friedman in the slow rotation non-Cowling
   limit.}  \label{fig:inertial}
\end{figure}

A graph showing the frequency of three such modes as a function of stellar
rotation rate is shown in figure \ref{fig:inertial}.  These are $m=2$ modes
whose leading contribution to the velocity field has harmonic index $l=4$.
Analytic fits to the radial eigenfunctions of these modes in uniform
density stars are given in table 3 of Lockitch \& Friedman.  As the latter
authors note, the eigenfunctions are very similar for $n=1$ polytropes, so
we have used these analytic functions to supply initial data to our code.
As in previous figures, the vertical axis plots the dimensionless mode
frequency as measured in the rotating frame, while the angular rotation
rate is given in units of $\sqrt{\pi G \rho_0}$.  The three horizontal
lines indicate the mode frequency in the slow rotation limit as calculated
by Lockitch \& Friedman (see table 6 of their paper).
In the limit of slow rotation we found the eigenfunctions were sufficiently
accurate to deposit approximately $99\%$ of their energy into a single
mode, with frequency very close to that computed by Lockitch \& Friedman,
confirming our identification of the mode.  The slight difference between
the frequencies is due to our having made the Cowling approximation, while
Lockitch \& Friedman have not.

\section{Conclusions}

In this paper we have presented results from a numerical code for time
evolutions of perturbations of 
rapidly and rigidly rotating Newtonian polytropes (within the
Cowling approximation).  Our code is second order convergent and runs stably
for many (i.e. hundreds) of stellar oscillations.  We have compared the
results of our evolutions with calculations of f- and r-modes in the
literature and found agreement within the errors expected for the Cowling
approximation.  We have also presented results for the inertial modes,
and (for the first time) extended the frequencies calculated by 
Lockitch \& Friedman into the
rapid rotation regime.

Even though we have provided some new results for oscillations in stars 
rotating near the break-up limit, this work is essentially a 
 ``proof of principle''. Our main aim was to illustrate the accuracy 
attainable with a time-evolution approach to the problem of 
rotating stars. In particular, the relative simplicity and 
reliability of the 
perturbative 2D approach makes it a valuable complement to 
fully 3D hydrodynamical simulations, such as the recent ones by 
Lindblom, Tohline and Vallisneri (2001). 

Having demonstrated the stability and accuracy of our code, we plan to use
it as a test-bed for the implementation of additional physical features,
which we will add in a modular fashion, one at a time.  Via such a gradual
escalation in complexity, we hope to arrive at a code capable of evolving
rather more realistic stars, incorporating (for instance) differential
rotation, gravitational radiation reaction, and non-linear effects.  Such
extensions are currently underway, and will be presented in due course.

\section*{ACKNOWLEDGEMENTS}

It is a pleasure to thank Shin Yoshida for supplying the r-mode data
used in this paper, Kostas Kokkotas for the f-mode data, and also Uli
Sperhake for help with numerical problems.  This work was supported by
PPARC grant PPA/G/1998/00606, and also by the EU Programme `Improving
the Human Research Potential and the Socio-Economic Knowledge Base'
(Research Training Network Contract HPRN-CT-2000-00137).

\end{document}